\documentclass[twocolumn,
english,aps,prl,superscriptaddress,showpacs,preprintnumbers]{revtex4}%
\usepackage[latin9]{inputenc}
\usepackage{units}
\usepackage{amsmath}
\usepackage{graphicx}
\usepackage{amssymb}
\usepackage{amsfonts}
\usepackage{verbatim}
\usepackage{natbib}
\providecommand{\U}[1]{\protect\rule{.1in}{.1in}}
\newcommand{\etac}{\ensuremath{\eta_{\mathrm{c}}}}

\begin{document}

\title{Optomechanical sideband cooling of a micromechanical oscillator 
close
to the quantum ground state}

\author{R. Rivi\`{e}re}
\thanks{These authors contributed equally to this work.}
\affiliation{Max Planck Institut f{\"u}r Quantenoptik, 85748 Garching, Germany}
\author{S. Del\'{e}glise$^{*}$}
\affiliation{Max Planck Institut f{\"u}r Quantenoptik, 85748 Garching, Germany}
\affiliation{\'{E}cole Polytechnique F\'{e}d\'{e}rale de Lausanne (EPFL), 1015 Lausanne, Switzerland}
\author{S. Weis$^{*}$}
\affiliation{Max Planck Institut f{\"u}r Quantenoptik, 85748 Garching, Germany}
\affiliation{\'{E}cole Polytechnique F\'{e}d\'{e}rale de Lausanne (EPFL), 1015 Lausanne, Switzerland}
\author{E. Gavartin}
\affiliation{\'{E}cole Polytechnique F\'{e}d\'{e}rale de Lausanne (EPFL), 1015 Lausanne, Switzerland}
\author{O. Arcizet}
\affiliation{Institut N\'eel, 38042 Grenoble, France}
\author{A. Schliesser}
\affiliation{Max Planck Institut f{\"u}r Quantenoptik, 85748 Garching, Germany}
\affiliation{\'{E}cole Polytechnique F\'{e}d\'{e}rale de Lausanne (EPFL), 1015 Lausanne, Switzerland}
\author{T.J. Kippenberg}
\email{tobias.kippenberg@epfl.ch}
\affiliation{Max Planck Institut f{\"u}r Quantenoptik, 85748 Garching, Germany}
\affiliation{\'{E}cole Polytechnique F\'{e}d\'{e}rale de Lausanne (EPFL), 1015 Lausanne, Switzerland}

\begin{abstract}
Cooling a mesoscopic mechanical oscillator to its quantum ground state is elementary for the preparation and control of low entropy quantum states of large scale objects.
Here, we pre-cool a  70-MHz micromechanical  silica oscillator to an occupancy below 200 quanta by thermalizing it with a 600-mK cold ${}^3$He gas. 
Two-level system induced damping via structural defect states is shown to be strongly reduced, and simultaneously serves as novel thermometry method to independently quantify excess heating due to a cooling laser.
We demonstrate that dynamical backaction sideband cooling can reduce the  average occupancy to $9\pm1$ quanta, implying that the mechanical oscillator can be found $(10\pm1) \%$ of the time in its quantum ground state.
\end{abstract}


\maketitle

\affiliation{Max Planck Institut f{\"u}r Quantenoptik, 85748 Garching, Germany}
\affiliation{\'{E}cole Polytechnique F\'{e}d\'{e}rale de Lausanne (EPFL), 1015 Lausanne, Switzerland}

The quantum regime of mechanical systems has received significant interest
over the past decade \cite{Schwab2005, Kippenberg2008, Marquardt2009, Favero2009a}. Mechanical systems cooled to the quantum ground state may allow probing quantum mechanical phenomena on an
unprecedentedly large scale, could enable quantum state preparation of
mechanical systems and have been proposed as an interface between
photons and stationary qubits. To achieve ground state cooling, two
challenges have to be met: first, most mechanical oscillators have vibrational
frequencies $\Omega_{\mathrm{m}}/2\pi<100\,\unit{MHz}$, such that low mode temperatures $T_{\mathrm{eff}}$ are required to achieve
$\hbar\Omega_{\mathrm{m}}> k_{\mathrm{B}}T_{\mathrm{eff}}$ ($\hbar$ is the reduced Planck constant and
$k_{\mathrm{B}}$ the Boltzman constant). Second,
quantum limited measurements of mechanical motion must be performed
at the level of the zero point motion, $x_{\mathrm{zpf}}=\sqrt{{\hbar
}/{2m_{\mathrm{eff}}\Omega_{m}}}$ in order to probe the state of the oscillator of mass $m_{\mathrm{eff}}$.

Recently, a piezomechanical oscillator has been cooled to the quantum regime
\cite{OConnell2010}. Due to its GHz resonance frequency, conventional
cryogenics could be employed for cooling, while it was probed using its
piezoelectrical coupling to a superconducting qubit. In contrast, cooling
schemes based on radiation pressure dynamical backaction as
proposed \cite{Braginskii1967, Dykman1978} and
recently demonstrated \cite{Schliesser2006, Arcizet2006a, Gigan2006} can be
applied to a much wider class of nano- and micromechanical oscillators. 
This optomechanical scheme is based on parametric coupling of an optical and mechanical resonance and simultaneously allows sensitive detection of mechanical motion. In analogy
to the case of trapped ions \cite{Leibfried2003}, dynamical backaction sideband
cooling \cite{Wilson-Rae2007, Marquardt2007, Bhattacharya2007a, Schliesser2008} can be used to
reach the quantum ground state.

Despite major progress, ground state cooling using this approach has remained
challenging, owing to insufficiently low starting temperatures or excess
heating in the optical domain \cite{Schliesser2009a, Park2009, Groblacher2009}%
, while microwave experiments have been impeded by the residual thermal
occupancy in the microwave cooling tone \cite{Rocheleau2010} or weak
optomechanical coupling \cite{Teufel2008}.
Here we demonstrate an experimental optomechanical setting that solves these challenges. 

\begin{figure}[bt]
\centering
\includegraphics[width=.8\linewidth]{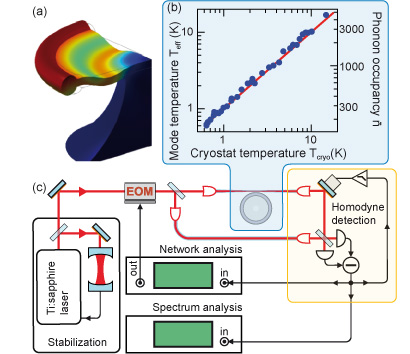}\caption{
Cooling a micromechanical oscillator. %
(a) High Q mechanical and optical modes are co-located in a silica microtoroid. The simulated displacement pattern of the mechanical
radial breathing mode (RBM) is shown; the optical whispering gallery mode (WGM) is confined to
the rim. (b) Thermalization of the RBM to the temperature of the ${}^3$He gas, with
the lowest achieved temperature corresponding to an occupancy of the RBM
below 200 quanta. (c) Optical setup used for displacement monitoring of the
mechanical mode, based on homodyne analysis of the light re-emerging from the
toroid's WGM (see text for detailed description).}%
\label{fig1}%
\end{figure}

We use silica toroidal resonators, which support whispering gallery modes (WGM) of ultrahigh
finesse co-located with a low loss mechanical radial breathing mode (RBM)
\cite{Kippenberg2005, Schliesser2010} and large mutual optomechanical coupling.
The devices used here have been
optimized for narrow optical linewidths $\kappa$, and moderately high
mechanical frequencies $\Omega_{\mathrm{m}}$, thereby operating deeply
in the resolved sideband regime ($\Omega_{\mathrm{m}}\approx2\pi\cdot70\,%
\operatorname{MHz}%
\gtrsim10\kappa$), while at the same time the pillar geometry was engineered
for low mechanical dissipation \cite{Schliesser2008,
Anetsberger2008} (Fig.\ 1).

For the cryogenic laser cooling experiments, we subject these samples directly to a ${}^{3}$He  gas evaporated from a reservoir of liquid ${}^{3}$He recondensed before each experimental run.
At a pressure of $\sim 0.7\,\unit{mbar}$, the gas provides a thermal bath at
a temperature of ca.\ $600\,\unit{mK}$.
However, it is essential to verify thermalization of the toroid to the exchange gas. To this end, a low-noise fiber laser
(wavelength $\lambda\approx1550\,%
\operatorname{nm}%
$) is coupled to a WGM using a fiber taper positioned in the near field of the mode via piezoelectric actuators (Attocube GmbH).
Using techniques described previously
\cite{Arcizet2009a, Gorodetsky2010}, the displacement fluctuations of the
RBM can be extracted and used to infer its noise temperature.
 As shown in Fig.\ \ref{fig1}b), it follows the temperature
of the helium gas down to temperatures of $600\,%
\operatorname{mK}%
$ for weak probing (i.e.\ $<1\,\unit{\mu W}$ input laser power).

For the optomechanical sideband cooling, we employed a
frequency-stabilized Ti:sapphire laser ($\lambda\approx 780\,\unit{nm}$), and a homodyne detection scheme
\cite{Schliesser2009a} for quantum-limited detection of mechanical
displacement fluctuations (Fig.\ \ref{fig1}c).
Importantly, for the high Fourier
frequencies of interest the Ti:sapphire laser is quantum limited in amplitude and
phase. Spectral analysis of this signal provides direct access to the
mechanical displacement spectrum, from which the mechanical damping and
resonance frequency can be derived. The spectra are calibrated in absolute
terms \cite{Schliesser2009a,Gorodetsky2010} by applying a known frequency
modulation at a fixed frequency close to the mechanical resonance frequency to
the laser using an electro-optic modulator (EOM).
 After the acquisition of each spectrum, the detuning of the
laser from the cavity resonance is determined by sweeping the modulation
frequency of the EOM and recording the demodulated homodyne signal with the
network analyzer \cite{Weis2010}.

Before studying radiation-pressure induced effects, we have carefully analyzed
the influence of the bath temperature on the RBM's properties. The vitreous
nature of silica leads to a strong temperature
dependence due to the presence of structural defects modeled as two-level-systems (TLS)
\cite{Arcizet2009a, Enss2005}. Relaxation of the TLS under excitation from an
acoustic wave modifies the complex mechanical susceptibility, leading to a
change in mechanical resonance frequency $\Omega_{\mathrm{m}}$ and a change of
the damping rate $\Gamma_{\mathrm{m}}=\Omega_{\mathrm{m}}/Q_{\mathrm{m}}$ with mechanical quality factor
 $Q_{\mathrm{m}}$.

\begin{figure}[ptbh]
\centering
\includegraphics[width=.8\linewidth]{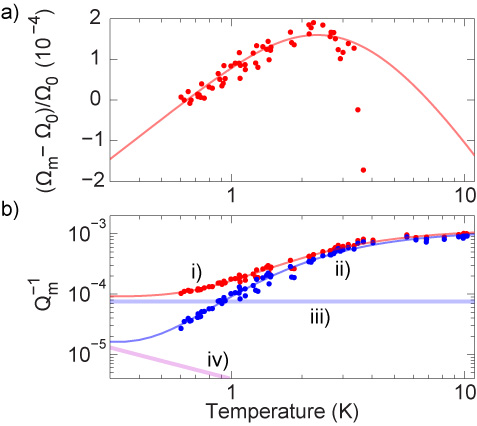}\caption{
TLS-induced change of the resonance frequency ($\Omega_{\mathrm{m}}$) (a) and inverse mechanical quality factor ($Q_{\mathrm{m}}^{-1}$) (b) of the radial breathing mode. Measured data (red points) agree well with the models
(red lines) described in eqs. (\ref{equationOmegaM}) and (\ref{equationGammaM}%
). Subtraction of the temperature-independent clamping damping (line
iii) yields the theoretically possible material-limited damping values (blue
points and line ii). At very low temperatures, damping by resonant
interaction with TLS (iv) would be dominant.
The model parameters are given in ref.\ \cite{si}.}%
\label{fig2}%
\end{figure}

Two different relaxation regimes have to be considered \cite{si} for sample
temperatures $T$ between $0.6\,%
\operatorname{K}%
$ to $3\,%
\operatorname{K}%
$: tunneling-assisted relaxation \cite{Phillips1987, Jaeckle1972}, and
single phonon resonant interaction \cite{Jaeckle1972}. Thermally activated
relaxation \cite{Vacher2005} dominates the frequency shift at temperatures
above $3\,%
\operatorname{K}%
$, but is negligible in the temperature range at which the laser cooling experiments are performed.
In the presence of tunneling relaxation (\textquotedblleft tun\textquotedblright) and resonant
interactions (\textquotedblleft res\textquotedblright) the mechanical
oscillator properties can be expressed as
\begin{align}
\Omega_{\mathrm{m}}(T) &  =\Omega_{\mathrm{m}}+\delta\Omega_{\mathrm{tun}%
}(T)+\delta\Omega_{\mathrm{res}}(T)\label{equationOmegaM}\\
\Gamma_{\mathrm{m}}(T)/\Omega_{\mathrm{m}} &  =Q_{\mathrm{m}}^{-1}%
(T)=Q_{\mathrm{cla}}^{-1}+Q_{\mathrm{tun}}^{-1}(T)+Q_{\mathrm{res}}%
^{-1}(T),\label{equationGammaM}%
\end{align}
where $\Omega_{\mathrm{m}}Q_{\mathrm{cla}}^{-1}$ is the damping rate due to the clamping of the resonator to the substrate,
dominating $\Gamma_{\mathrm{m}}$ at room temperature. The respective temperature
dependencies in the relevant regimes of TLS damping are detailed in \cite{si}.
For the lowest temperatures of $600\,\unit{mK}$, $Q_{\mathrm{m}}$ reaches $\sim10^4$ sufficient
to enable ground state cooling since $Q_{\mathrm{m}}/\bar{n}_{\mathrm{i}}\gg1$ and
$\bar n_\mathrm{i} \Omega_{\mathrm{m}}/Q_{\mathrm{m}}\ll\kappa$ ($\bar{n}_{\mathrm{i}}$ is the initial occupancy) \cite{Dobrindt2008}.
Moreover, the well-understood temperature dependence of the TLS-induced effects [Eqs.\ (\ref{equationOmegaM})-(\ref{equationGammaM})] enables its use as a ``thermometer'' of the sample temperature $T$ after a calibration measurement as shown in Fig.\ \ref{fig2} has been performed once.
Importantly, this method can reveal excess heating independent of the RBM's noise temperature.

\begin{figure}[ptbh]
\centering
\includegraphics[width=.8\linewidth]{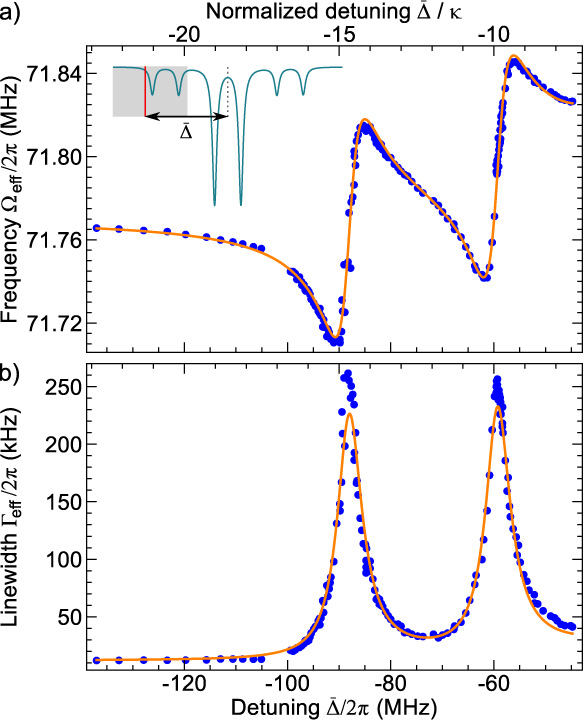}\caption{Effective resonance frequency (a)
and linewidth (b) of the RBM when a $2\,\operatorname{mW}$-power laser is
tuned through the lower mechanical sideband of the split optical mode (inset).
Blue points are measured data extracted from the recorded spectra of thermally induced
mechanical displacement fluctuations, solid lines are a coupled fit
using the model of eqs.\ (\ref{e:chi})-(\ref{e:Oeff}), taking into account
heating of the cavity due to absorbed stray and intracavity light, modifying
the mechanical properties via the temperature dependence of the TLS. }%
\label{fig3}%
\end{figure}

We next studied optomechanical cooling \cite{Schliesser2006, Arcizet2006a, Gigan2006} by
performing a series of experiments in which mechanical displacement noise spectra were
recorded while varying the laser detuning $\bar{\Delta}\equiv\omega_{\mathrm{l}%
}-\bar{\omega}_{\mathrm{c}}$.
Here,  $\omega_{\mathrm{l}}$ is the laser's (angular) frequency and  $\bar{\omega}_{\mathrm{c}}$
the WGM frequency, taking temperature and static radiation-pressure induced shifts into account.
The mechanical mode's
frequency and damping shows a strong detuning dependence (Fig.\ \ref{fig3})
owing to the dynamic in-phase and quadrature response of the radiation pressure force
with respect to the mechanical motion.

To accurately model radiation-pressure induced dynamical backaction \cite{Braginskii1967} for
the present microresonators, we have to additionally take into account that backscattering
of light can couple WGMs with opposite circulation
sense \cite{si,Weiss1995,Kippenberg2002}. The mutual coupling of the clockwise
($a_{\mathrm{cw}}$) and counterclockwise ($a_{\mathrm{ccw}}$) orbiting modes
lifts the degeneracy leading to new eigenmodes of the system, i.e.
$a_{+}\equiv(a_{\mathrm{ccw}}+a_{\mathrm{cw}})/\sqrt{2}$ and $a_{-}%
\equiv(a_{\mathrm{ccw}}-a_{\mathrm{cw}})/\sqrt{2}$, where the new
eigenfrequencies $\bar{\omega}_{\pm}=\bar{\omega}_{\mathrm{c}}\mp\gamma/2$ are split by the mutual
coupling rate~$\gamma$. During a detuning series as reported here, both modes
are populated by the driving field $s_{\mathrm{in}}$ with a mean field
$\bar{a}_{\pm}=\sqrt{\kappa_{\mathrm{ex}}/2}\,L_{\pm}(\bar{\Delta
})s_{\mathrm{in}}$, where $P_{\mathrm{in}}=|s_{\mathrm{in}}|^{2}\hbar
\omega_{\mathrm{l}}$ is the driving laser power, $\kappa_{\mathrm{ex}}$ the
coupling rate to the fiber taper, $|\bar{a}_{\pm}|^{2}$ the mean photon
population in the new eigenmodes and $L_{\pm}(\bar{\Delta})\equiv\left(
-i(\bar{\Delta}\pm\gamma/2)+\kappa/2\right)  ^{-1}$ the modes' Lorentzian
response.

In the context of cavity optomechanics, it is important to realize that
three-mode interactions \cite{Braginsky2001} can be neglected in the present
configuration \cite{si}.
The radiation pressure forces induced by the light in these modes can
therefore simply be added. The usual linearization procedure \cite{Fabre1994,Mancini1994} then
yields an inverse effective mechanical susceptibility of
\begin{equation}
\left(  \chi_{\mathrm{eff}}(\Omega)\right)  ^{-1}=m_{\mathrm{eff}}\left(
\Omega_{\mathrm{m}}^{2}-\Omega^{2}-i\Gamma_{\mathrm{m}}\Omega-i\Omega
_{\mathrm{m}}f(\Omega)\right)  \label{e:chi}%
\end{equation}
modified by dynamical backaction according to
\begin{equation}
f(\Omega)=2g_{0}^{2}\sum_{\sigma=\pm}|\bar{a}_{\sigma}|^{2}\left(  L_{\sigma
}(\bar{\Delta}+\Omega)-(L_{\sigma}(\bar{\Delta}-\Omega))^{\ast}\right)
\end{equation}
with the vacuum optomechanical coupling rate
\cite{Gorodetsky2010} $g_{0}\equiv G x_{\mathrm{zpf}}$ and $G=\mathrm{d} \omega_{\mathrm{c}}/\mathrm{d} x$.
 For moderate driving powers \cite{Dobrindt2008}, the susceptibility of the mechanical mode is the one of a harmonic oscillator with effective
damping rate and resonance frequency of
\begin{align}
\Gamma_{\mathrm{eff}} &  \approx\Gamma_{\mathrm{m}}(T)+\mathrm{Re}\left[
f(\Omega_{\mathrm{m}})\right]  \\
\Omega_{\mathrm{eff}} &  \approx\Omega_{\mathrm{m}}(T)+\mathrm{Im}\left[
f(\Omega_{\mathrm{m}})\right]  /2.\label{e:Oeff}%
\end{align}

For the samples studied in the following, a coupling rate of $|g_{0}|\approx 2\pi\times (1.2\pm0.2)\,\unit{kHz}$ is determined from the coupling parameter $|G|= \omega_{\mathrm{c}}/R \approx 2 \pi \times 16 \,\unit{GHz/nm}$ and effective masses $m_{\mathrm{eff}}=20\pm5\,\unit{ng}$.
Figure \ref{fig3} shows the results of a detuning series, which was taken with an input laser power of $2\,\unit{mW}$, with the temperature of the ${}^3\mathrm{He}$ gas stabilized to $T_{\mathrm{cryo}}=850\,\unit{mK}$ at a pressure of $2.8\,\unit{mbar}$.
The excellent stability of both the laser and the cryogenic coupling setup allowed us to perform the series without active stabilization of the laser detuning $\bar{ \Delta}$ and the coupling $\kappa_{\mathrm{ex}}$, which is determined by the sub-micrometer gap between the coupling taper and the toroid.

The coupled fit of the data using the model of eqs.\ (\ref{e:chi})-(\ref{e:Oeff}) is enabled by the precise calibration of the laser detuning by sweeping the modulation frequency \cite{Weis2010}.
We choose to adjust the parameters of the fit primarily (relative weight $0.9$)  to the optical spring effect, since the mechanical resonance frequency can be extracted from the spectra with higher accuracy than the damping rate.
The obtained fit parameters $\kappa$, $\gamma$, and $s_{\mathrm{in}}$ are found to be in good agreement with independent results deduced from the frequency modulation measurement ($\kappa\approx 2\pi\times 6\,\unit{MHz}$, $\gamma\approx2\pi \times 30\,\unit{MHz}$) and the measured laser power.

The excellent quality of the fit, together with the measured temperature dependence of the TLS effects on the mechanical mode, furthermore allows us to extract the  temperature $T$ of the sample.
Importantly, for large detunings $|\bar{\Delta}| \gg \kappa$, the TLS thermometer reveals an increase of the sample temperature by $\Delta T_{\mathrm{stray}}\approx220\,\unit{mK}$ corresponding to $Q_{\mathrm{m}}(T)=5970$, which we attribute to heating induced by absorbed stray light scattered from defects on the fiber taper, which were observed to aggregate upon its production.

As the laser is tuned closer to resonance, more light is coupled into the WGM and
\begin{equation}
  T\approx T_{\mathrm{cryo}}+\Delta T_{\mathrm{stray}}+\Delta T_{\mathrm{WGM}},
\end{equation}
where $\Delta T_{\mathrm{WGM}}=\beta \kappa_{\mathrm{abs}} (|a_{+}|^2+|a_{-}|^2) \hbar \omega_{\mathrm{l}}$ denotes the increase in temperature following the cavity's double-Lorentizan absorption profile, $\kappa_{\mathrm{abs}}\lesssim \kappa-\kappa_{\mathrm{ex}}$ is the photon absorption rate and $\beta$ the temperature increase per absorbed power.
Operating deeply in the resolved sideband regime  \cite{Schliesser2008}, only little optical power ($\sim\kappa_{\mathrm{abs}} |a_{+}|^2 \hbar\omega_{\mathrm{l}}\lesssim P_{\mathrm{in}}/1300$) can be absorbed in the cavity when $\bar{\Delta}=-\Omega_{\mathrm{m}}-\gamma/2$, leading to a modest temperature increase of $\Delta T_{\mathrm{WGM}}\approx 70\,\unit{mK}$.

Importantly, we can test the consistency of the derived detuning-dependent quantities--- 
$T$%
, %
$\Gamma_{\mathrm{m}}(T)$%
, and  
$\Gamma_{\mathrm{eff}}$%
---by comparing the expected effective temperature of the RBM due to optomechanical cooling, i.e. $T_{\mathrm{eff}}=T\cdot \Gamma_{\mathrm{m}}(T)/\Gamma_{\mathrm{eff}}$, with the effective temperature derived from noise thermometry via integration of the calibrated noise spectra \cite{Schliesser2009a}.
Figure 4a) shows this comparison for the detuning series discussed above.
Using the model of eqs.\ (\ref{equationGammaM})-(\ref{e:Oeff}) adjusted to the data of Fig.\ \ref{fig3}, we obtain good agreement for the effective temperatures obtained in both ways.
To achieve this level of agreement, it is necessary to take into account the optomechanical de-amplification of the laser phase modulation used for calibrating the mechanical fluctuation spectra in absolute terms:
as was shown in a recent study \cite{Verlot2010a}, the transduction of a phase modulation of depth $\delta\varphi$ at a frequency $\Omega_{\mathrm{mod}} / 2 \pi$ is modified by a factor $| \chi_{\mathrm{eff}}(\Omega_{\mathrm{mod}}) / \chi_{\mathrm{m}}(\Omega_{\mathrm{mod}}) |$ in the presence of dynamical backaction, where $\chi_{\mathrm{m}}(\Omega)$ is the bare mechanical susceptibility.

Figure 4b) shows the same comparison for a cooling run at a high laser power ($4\,\unit{mW}$), for which we observe slightly increased heating by $\Delta T_{\mathrm{stray}}\approx400\,\unit{mK}$, while additional heating by $\Delta T_{\mathrm{WGM}}$ could not be discerned in this measurement.
In spite of the reduced mechanical quality factor
$Q_{\mathrm{m}}(T_{\mathrm{cryo}}+\Delta T_{\mathrm{stray}})=4880$, the lowest extracted occupancy is $\bar{n}=10$ according to the detuning series fit.
The lowest inferred \emph{noise} temperature of a single measurement is even slightly lower, corresponding to $\bar{n}=9\pm1$, where the uncertainty is dominated by systematic errors, which we estimate from the deviations of the effective temperature derived in the two independent ways described above.
Note that this occupancy implies already a probability of $P(n=0)=(1+\bar{n})^{-1}=(10\pm1) \%$ to find the oscillator in its quantum ground state.
From this measurement, we can also extract a total force noise spectral density of $S_{FF}=(8\pm2 \,\unit{fN}/\sqrt{\unit{Hz}})^2$ driving the oscillator  to this occupancy, corresponding to a imprecision-backaction product
 \cite{Schliesser2009a, si} of $\sqrt{S_{xx}^{\mathrm{im}}S_{FF}^{\mathrm{}}}= (49\pm8) \hbar/2$  if the entire present force noise is (conservatively) considered as measurement backaction.
Indeed, we estimate  about $40\%$ of the force noise to originate from the Langevin force at the temperature of the cryostat, and $60\%$ from the excess Langevin force caused by the laser-induced heating by $\Delta T_{\mathrm{stray}}$ \cite{si}.
Due to the resolved-sideband operation, force noise due to quantum backaction (as yet only observed on cold atomic gases \cite{Murch2008})  is expected to be nearly two orders of magnitude weaker and therefore negligible.

\begin{figure}[tbhp]
\centering
\includegraphics[width=.8\linewidth]{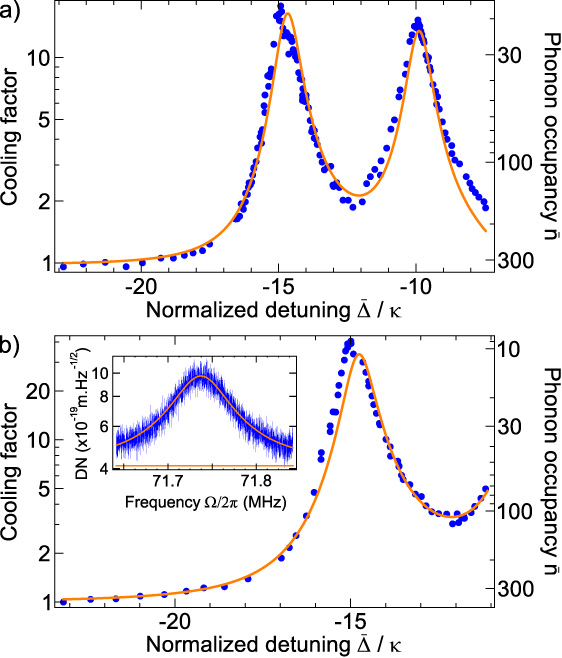}
\caption{Cooling factor $(T_{\mathrm{cryo}}+\Delta T_{\mathrm{stray}})/T_{\mathrm{eff}}$ and phonon occupancy number of the RBM versus laser detuning $\bar{\Delta}$ for (a) $P_{\mathrm{in}}=2\,\unit{mW}$ and (b) $P_{\mathrm{in}}=4\,\unit{mW}$.
Points are phonon occupancies derived from the measured noise temperature, while solid lines correspond to the occupancies expected from the dependency of sample temperature and (intrinsic and effective) damping extracted from the detuning series as shown in Fig.\ \ref{fig3}.
A minimum phonon number of $\bar{n}=9\pm1$ is obtained.
The inset shows a mechanical displacement noise (DN) spectrum at the optimum detuning ($\bar{\Delta}=-\Omega_{\mathrm{m}}-\gamma/2$), illustrating the signal-to-noise ratio achieved despite the low occupancy.
}
\label{fig4}
\end{figure}

It is realistic to significantly reduce the occupancy by higher cooling powers and improvements of $g_0^2/\Gamma_{\mathrm{m}}$, which can be achieved by engineering mechanical modes \cite{Anetsberger2008} for smaller mass and lower damping.
For occupancies $\bar{n}\lesssim1$, we anticipate that individually resolved anti-Stokes and Stokes sidebands \cite{Wilson-Rae2007, Schliesser2008} of an independent readout laser will display a measurable asymmetry of $\bar{n}/(\bar{n}+1)$ arising from the non-zero commutator of the ladder operators describing the mechanical harmonic oscillator in quantum mechanical terms. 

%
%


\newpage
\widetext
\newpage

\renewcommand{\thefigure}{\textbf{S}\arabic{figure}}
\renewcommand{\theequation}{$\mathrm{S} $\arabic{equation}}
\setcounter {figure} {0}
\setcounter {equation} {0}
\begin{center}
\large{\textbf{
Supplementary information - Optomechanical sideband cooling of a micromechanical oscillator 
close
to the quantum ground state}}
\end{center}
\vspace{.2in}

\section{Two-level systems}

Tunneling systems in $\mbox{SiO}_{{2}}$ play an important role in cryogenic operation of silica mechanical oscillators.
They lead to a temperature dependent frequency shift (via a change in speed of sound) of the considered mechanical mode and temperature dependent mean free path of phonons in $\mbox{SiO}_{{2}}$ affecting the mechanical quality factor.
We will discuss these effects in the following and give the relevant formulae which have been used to fit the data in the main part of the manuscript.
An extensive study of TLS effects can be found in \cite{SIEnss2005,SIJaeckle1972}.

\begin{figure}[bth]
\centering{}\includegraphics[width=0.4\textwidth]{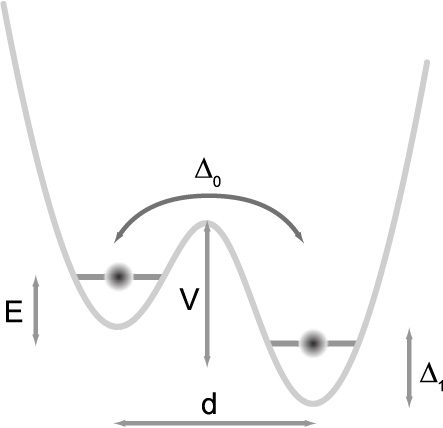}\caption{Double well potential with relevant levels and naming convention\label{fig:doublewell}}
\end{figure}

As first considered by L.~Pauling in 1930 \cite{SIPauling}, tunneling of atoms occurs in solids with a certain degree of disorder, where in the local environment of an atomic site, several potential minima exist.
This can be the case in the vicinity of defects in crystals or, more frequently, in amorphous materials.
For amorphous solids at low temperatures, the tunneling dynamics can be well captured in a simple model consisting of an ensemble of two level systems (TLS) each of which is described by  a  generic double-well potential (Fig. \ref{fig:doublewell}).
This potential is parametrized only by the barrier height $V$, the initial energy asymmetry $\Delta_\mathrm{1}$ and the spatial separation between the two potential minima $d$.
A tunneling coupling strength
\begin{equation}
  \Delta_{0}=\hbar\Omega_{0}e^{-\lambda}
\end{equation}
with the intrinsic oscillation frequency $\Omega_{0}$ within the individual atomic sites can then be deduced, with the tunneling parameter
\begin{equation}
  \lambda\approx\sqrt{\frac{2 m V}{\hbar^2}}\frac{d}{2}
\end{equation}
depending on the atomic mass $m$.
Due to this tunnel coupling, the new eigenmodes of the coupled system exhibit an energy splitting of
\begin{equation}
  E=\sqrt{\Delta_\mathrm{1}^{2}+\Delta_{0}^{2}}.
\end{equation}

Phonons couple to TLS via their strain field that leads to a deformation of the TLS potential (notably leading to a change in the barrier height V).
As a consequence, the TLS are driven out of thermal equilibrium and relaxation processes will exchange energy with the heat bath.
Transitions between the two energy levels are induced by several distinct processes that become dominant in different temperature regimes.

\begin{itemize}
\item At very low temperatures ($T\lesssim E/k_\mathrm{B}$), the density of thermal phonons is low, such that relaxation processes do only play a minor role.
Here a significant population imbalance between lower and excited state exists, and the most efficient transition mechanism is resonant absorption of phonons of a frequency $\Omega_\mathrm{m}=E/\hbar$.
This mechanism shows---as in the case of other two level systems---a saturation behavior \cite{SIHunklinger72}.

\item At temperatures $T\gtrsim E/k_\mathrm{B}$ (typically a few Kelvin), the number of thermal phonons has increased to a level at which Raman processes involving a tunneling through the barrier become predominant.
It is mainly this temperature range that will be relevant for the description of the phenomena seen in the present cooling experiment.

\item At even higher temperatures thermally activated relaxation dominates.
In this case multi-phonon processes with an excitation across the barrier take place.
\end{itemize}

\subsection{Relaxation contribution}
If relaxation is the dominant process, the general expression for the mean free path of a phonon of frequency $\Omega_\mathrm{m}$ is given
by \cite{SIEnss2005,SIJaeckle1972}
\begin{equation}
  l^{-1}(T)=
  \frac{1}{\rho c_\mathrm{s}^{3}}
  \iint
  \left(-\frac{\partial n_{0}}{\partial E}\right)
  4B^{2}\frac{\Delta_\mathrm{1}^{2}}{E^{2}}\frac{\Omega_\mathrm{m}^{2}\tau}{1+\Omega_\mathrm{m}^{2}\tau^{2}}
  \bar{P}(\Delta_\mathrm{1},\lambda)\, \mathrm{d}\Delta_\mathrm{1}\, \mathrm{d}\lambda.
\end{equation}
The integration is performed on all TLS that can interact with the phonon.
Here, $\bar{P}(\Delta_\mathrm{1},\lambda)$ is the volume density of TLS with energy asymmetry between $\Delta_\mathrm{1}$ and $\Delta_\mathrm{1}+\mathrm{d}\Delta_\mathrm{1}$ and tunnel parameter between $\lambda$ and $\lambda+\mathrm{d}\lambda$,
\begin{equation}
  n_{0}=\frac{1}{e^{E/k_{B}T}+1}
\end{equation}
is the thermal equilibrium Boltzmann repartition function, $c_\mathrm{s}$ the speed of sound, $\rho$ the mass density of the solid, $\tau$ the relaxation time of the individual TLS and $B$ the coefficient linking a deformation $\delta e$ to a change of $E$ via $\delta E=2B(\Delta_\mathrm{1}/E)\delta e$.

A mechanical quality factor of
\begin{equation}
  Q_{\mathrm{m}}^{-1}(T)=\frac{c_\mathrm{s}l^{-1}(T)}{\Omega_\mathrm{m}}
\end{equation}
can then be deduced.
For the corresponding relative change in the speed of sound (i.e.\ frequency shift of a mechanical resonance) one obtains from the Kramers-Kronig
relations
\begin{equation}
 {\delta\Omega_\mathrm{m}}{}(T)=
  -\frac{\Omega_\mathrm{m}}{2\rho c_\mathrm{s}^{2}}
  \iint
    \left(-\frac{\partial n_{0}}{\partial E}\right)
  4B^{2}\frac{\Delta_\mathrm{1}^{2}}{E^{2}}\frac{1}{1+\Omega_\mathrm{m}^{2}\tau^{2}}
  \bar{P}(\Delta_\mathrm{1},\lambda)\, \mathrm{d}\Delta_\mathrm{1}\, \mathrm{d}\lambda.
\end{equation}

\subsubsection{Tunneling-assisted relaxation}

Within the framework of the so-called tunneling model \cite{SIEnss2005,SIJaeckle1972} the relaxation time is given by
\begin{align}
  \tau&=\tau_\mathrm{m}\frac{E^{2}}{\Delta_{0}^{2}}
  \intertext{with the maximum relaxation rate}
  \tau_\mathrm{m}^{-1}&=\frac{3}{c_\mathrm{s}^{5}}\frac{B^{2}}{2\pi\rho\hbar^{4}}E^{3}\coth\left(\frac{E}{2k_\mathrm{B}T}\right).
\end{align}
Parametrizing the integrals in terms of the energy splitting $E$ and the parameter $u=\tau^{-1}/\tau_\mathrm{m}^{-1}$ yields \cite{SIEnss2005,SIJaeckle1972}
\begin{align}
  Q^{-1}_\mathrm{tun}(T)&=
    \frac{2\bar{P}B^{2}}{\rho c_\mathrm{s}^{2}}
    \int_{0}^{\infty}
    \left(-\frac{\partial n_{0}}{\partial E}\right)
      \Omega_\mathrm{m}\tau_\mathrm{m}
      \int_{0}^{1}
        \frac{\sqrt{1-u}}{u^{2}+\Omega_\mathrm{m}^{2}\tau_\mathrm{m}^{2}}
      \,\mathrm{d}u
    \,\mathrm{d}E
  \intertext{and}
 {\delta\Omega_\mathrm{tun}}{}(T)
  &=-\frac{\Omega_\mathrm{m}\bar{P}B^{2}}{\rho c_\mathrm{s}^{2}}
  \int_{0}^{\infty}
    \left(-\frac{\partial n_{0}}{\partial E}\right)
    \int_{0}^{1}
      \frac{u\sqrt{1-u}}{u^{2}+\Omega_\mathrm{m}^{2}\tau_\mathrm{m}^{2}}
    \,\mathrm{d}u
  \,\mathrm{d}E,
\end{align}
where it is assumed that the density $\bar P(E,\lambda)=\bar P$ is constant, which is consistent with experiments.
A prominent feature in this regime is a plateau of the quality factors for temperatures of a few Kelvins with $Q$ values of
\begin{equation}
Q_{\mbox{plateau}}^{-1}=\frac{\pi}{2}\frac{\bar{P}B^{2}}{\rho c_\mathrm{s}^{2}}.
\end{equation}

\subsubsection{Thermally activated relaxation}
At higher temperature the rate is given by the Arrhenius law and only depends on the energy barrier height,
\begin{equation}
  \tau_\mathrm{th}^{-1}=\tau_{0}^{-1}e^{-V/k_\mathrm{B}T},
\end{equation}
where $\tau_{0}$ represents the period of oscillation in individual
wells \cite{SIVacher2005,SIEnss2005}.

\subsection{Resonant processes}

For resonant interaction between phonons and TLS, it can be shown that \cite{SIEnss2005,SIJaeckle1972}
\begin{align}
  Q_\mathrm{res}^{-1}(T)
  &=\frac{\pi\bar{P}B^{2}}{\rho c_\mathrm{s}^{2}}\tanh\left(\frac{\hbar\Omega_\mathrm{m}}{2k_{B}T}\right)\\
  \delta\Omega_\mathrm{res}(T)
  &=\frac{\Omega_\mathrm{m} \bar{P}B^{2}}{\rho c_\mathrm{s}^{2}}\ln \left(\frac{T}{T_{0}}\right),
\end{align}
where $T_{0}$ is a reference temperature.
While resonant processes do not significantly contribute to the mechanical quality
factors in our experiment, the frequency shift is dominated by resonant processes.

\subsection{Fitting Parameters for Figure 2}
The curves shown in figure 2 of the main manuscript have been fitted with the equations given in the previous sections.
For the frequency shift the sum of the tunneling relaxation and the resonant contribution has been taken into account.
The latter dominates this effect up to about $T=2\,\unit{K}$.
The contribution of thermally activated relaxation has been omitted since it doesn't contribute significantly in the considered temperature range.
Fitting of the $Q$-dependency has been done using the sum of tunneling relaxation, resonant contribution and a constant offset accounting for the clamping losses ($Q_{\mathrm{cla}}^{-1}$), i.e.\ loss of acoustic energy due to leaking into the substrate for this particular toroid.  Here the resonant contribution plays a minor role.

For the curves shown in Fig.\ 2 of the main manuscript, we used the material parameters
\begin{align*}
  c_{\mathrm{s}}&=5800 \,\unit{m/s}\\
  \rho&=2330 \,\unit{kg/m}^{3},
\intertext{the measured resonance frequency}
  \Omega_{\mathrm{m}}&=2 \pi\times 76.3 \,\unit{MHz},
\intertext{as well as the adjusted parameters}
  B&=1.1 \times 10^{-19}\,\unit{J}\\
  \bar{P}_{Q_{\mathrm{m}}}&=2.5 \times 10^{45}\,\unit{m}^{-3}\\
  \bar{P}_{\Omega_{\mathrm{m}}}&=4.6 \times 10^{45}\,\unit{m}^{-3}.
\end{align*}
For the fitting of the two curves (mechanical quality factor, resonance frequency shift) two different values for $\bar{P}$ had to be used.
Given that the two traces are governed by two different regimes, small differences in the density of contributing TLS to the two effects seem to be justified.
The literature \cite{SIPohl2002} value of the dimensionless parameter
$C=\bar{P} B^2/(\rho c_s^2)=3.0 10^{-4}$ shows a reasonable  agreement with the parameters of the resonance frequency  ($C_{\Omega_{\mathrm{m}}}=7.1 10^{-4}$) and damping ($C_{Q_{\mathrm{m}}}=3.9 10^{-4}$) fits.

\section{Dynamical backaction in the presence of mode splitting}

In the framework of coupled-mode theory \cite{SIHaus1984}, the two coupled counterpropagating modes \cite{SIWeiss1995,  SIKippenberg2002} in a WGM resonator can be described by the equations of motion (in a frame rotating at the laser frequency)
\begin{align}
\dot{a}_{\mathrm{ccw}}(t) &= (i (\Delta-G x(t)) - \kappa /2) a_{\mathrm{ccw}}(t)  + i
\frac{\gamma}{2} a_{\mathrm{cw}}(t)+\sqrt{\etac \kappa} s_\mathrm{in}(t)\\
\dot{a}_{\mathrm{cw}}(t) &= (i (\Delta-G x(t)) - \kappa /2)
a_{\mathrm{cw}}(t)  + i
\frac{\gamma}{2} a_{\mathrm{ccw}}(t).
\end{align}
Here $\etac$ describes the coupling parameters defined via $\etac = \frac{\kappa_\mathrm{ex}}{\kappa_\mathrm{ex} + \kappa_\mathrm{0}}$, where $\kappa_\mathrm{ex}$ describes the output coupling rate, whereas $\kappa_\mathrm{0}$ denotes the intrinsic loss rate of the cavity.

The fields in the system's new eigenmodes
\begin{align}
 a_+ &= (a_{\mathrm{ccw}} + a_{\mathrm{cw}})/\sqrt{2} \\
a_- &=(a_{\mathrm{ccw}} - a_{\mathrm{cw}})/\sqrt{2}
\intertext{exert a radiation pressure force of }
 F_{\mathrm{rp}}&=-{\hbar G}\left(|a_+(t)|^2 + |a_-(t)|^2\right),
\intertext{since the spatial shape of cross-term $2\mathrm{Re}(a_+(t) a_-^*(t))$ has an azimuthal dependence $\propto \cos(m\varphi)\sin(m\varphi)$ ($m$ is the angular mode number), averaging to zero when projected on the azimuthally symmetric RBM.
The coupled optomechanical equations of motion can therefore be written as}  
\dot{a}_+(t) &=  \left(i \left(\Delta -G x(t) +\frac{\gamma}{2}\right)   - \frac{\kappa}{2} \right) a_+(t)  + \sqrt{\frac{\etac \kappa}{2}} s_\mathrm{in}(t)\\
\dot{a}_-(t)  &=  \left(i \left(\Delta - G x(t) - \frac{\gamma}{2}\right) - \frac{\kappa}{2} \right) a_-(t)   + \sqrt{\frac{\etac \kappa}{2}} s_\mathrm{in}(t)\\
m_\mathrm{eff} \left(\ddot x(t) + \Gamma_ \mathrm {m} \dot x(t) + \Omega_ \mathrm {m}^2 x(t) \right) &=
-{\hbar G}\left(|a_+(t)|^2 + |a_-(t)|^2\right)+ \delta F(t),
\end{align}
where $\delta F(t)$ is an external force, e.\ g.\ the thermal Langevin force.

We then apply the usual linearization 
\begin{align}
 a_{\pm}(t)&=\bar{a}_{\pm}+\delta a_{\pm}(t)\\
 x(t)&=\bar{x}+\delta x(t)
 \intertext{assuming $|\bar{a}_{\pm}|\gg|\delta a_{\pm}(t)|$ and $|\bar{x}|\gg|\delta x(t)|$.
 For the large mean occupancy of the modes and the mean mechanical displacement, we then obtain } 
  \bar a_+&=\frac{\sqrt {\etac{\kappa}/{2}}\,\bar s_\mathrm{in}}{-i( \bar{\Delta}+\gamma/2)+{\kappa}/{2}}
  =:\sqrt {\etac{\kappa}/{2}}\,L_+(\bar{ \Delta})\,\bar s_\mathrm{in}\\
  \bar a_-&=\frac{\sqrt {\etac \kappa/2}\,\bar s_\mathrm{in}}{-i( \bar{\Delta}-\gamma/2)+\kappa/2}
    =:\sqrt {\etac{\kappa}/{2}}\,L_-(\bar{ \Delta})\,\bar s_\mathrm{in}\\
  \bar x&= -\frac{\hbar G}{m_\mathrm{eff} \Omega_\mathrm{m}^2} \left(|\bar a_+|^2 + |\bar a_-|^2\right).
  \intertext{The average displacement $\bar {x}$ induces a small static frequency shift, as does the (usually dominant) static shift due to absorption-induced heating \cite{SICarmon2004a}, which are both absorbed into the  mean detuning
  $\bar \Delta=\omega_{\mathrm{l}}-\left(\omega_{\mathrm{c}}(T) + G \bar x\right)$.
 One then obtains the equations of motion of small fluctuations,
   }
\delta \dot{a}_+(t) &=  \left(i\left(\bar \Delta +\frac{\gamma}{2}\right)- \frac{\kappa}{2} \right) \delta a_+(t) -i G \bar a_+ \delta x(t)\\
\delta \dot{a}_-(t) &=  \left(i\left(\bar \Delta -\frac{\gamma}{2}\right)- \frac{\kappa}{2} \right) \delta a_-(t) -i G \bar a_- \delta x(t)\\
m_\mathrm{eff} \left(\delta\ddot x(t) + \Gamma_ \mathrm {m}\delta\dot x(t) + \Omega_ \mathrm {m}^2 \delta x(t) \right) &=
 -\hbar G \left(\bar a_+^* \delta a_+(t)+ \bar a_+ \left(\delta a_+(t)\right)^*+\bar a_-^* \delta a_-(t)+\bar a_- \left(\delta a_-(t)\right)^*\right)+ \delta F(t).
\end{align}
Fourier transformation gives
\begin{align}
\delta {a}_+(\Omega) &= \frac{ -i G \bar a_+ \delta x(\Omega)}{ -i\left(\bar \Delta +{\gamma}/{2}+\Omega\right)+ {\kappa}/{2}}
=-i G \bar a_+ \,L_+(\bar \Delta+\Omega)\,\delta x(\Omega)\\
\delta {a}_-(\Omega) &= \frac{ -i G \bar a_- \delta x(\Omega)}{ -i\left(\bar \Delta -{\gamma}/{2}+\Omega\right)+ {\kappa}/{2}}
=-i G \bar a_- \,L_-(\bar \Delta+\Omega)\,\delta x(\Omega)\\
\delta x(\Omega)/\chi_\mathrm{m}(\Omega) &=
 -\hbar G \left(\bar{a}_+^* \delta a_+(+ \Omega)+ \bar a_+ \left(\delta a_+(-\Omega)\right)^*+\bar a_-^* \delta a_-(+ \Omega)+\bar a_- \left(\delta a_-(-\Omega)\right)^*\right)+ \delta F(\Omega).
 \end{align}
 With
 \begin{align}
  \chi_{\mathrm{m}}(\Omega)&=\frac{1}{m_\mathrm{eff} \left(-\Omega^2 -i\Omega \Gamma_ \mathrm {m} + \Omega_ \mathrm {m}^2  \right) }
  \intertext{Solving equations (S32 - S34) for $\delta x$ yields}
\delta x(\Omega)&=
\frac{ \delta F(\Omega)}
{
1/\chi_{\mathrm{m}}(\Omega)- i \hbar G^2
 \left(
 	|\bar a_+|^2 \left(L_+(\bar\Delta+ \Omega)- \left(L_+(\bar\Delta-\Omega)\right)^*\right)+
  	|\bar a_-|^2  \left(L_-(\bar\Delta+ \Omega)- \left(L_-(\bar\Delta-\Omega)\right)^*\right)
 \right)}
 \intertext{so that we can write}
 \frac{1}{\chi_\mathrm{eff}(\Omega)}&=
 \frac{1}{\chi_\mathrm{m}(\Omega)}- i \hbar G^2
 \left(
 	|\bar a_+|^2 \left(L_+(\bar\Delta+ \Omega)- \left(L_+(\bar\Delta-\Omega)\right)^*\right)+
  	|\bar a_-|^2  \left(L_-(\bar\Delta+ \Omega)- \left(L_-(\bar\Delta-\Omega)\right)^*\right)
 \right)
 \intertext{and, in the regime of weak optomechanical coupling \cite{SIDobrindt2008}}
 \Gamma_\mathrm{eff}&\approx
 	\Gamma_\mathrm{m}
	+ 2 x_\mathrm{zpf}^2 G^2 \mathrm{Re} \left(
 	|\bar a_+|^2 \left(L_+(\bar\Delta+ \Omega)- \left(L_+(\bar\Delta-\Omega)\right)^*\right)+
  	|\bar a_-|^2  \left(L_-(\bar\Delta+ \Omega)- \left(L_-(\bar\Delta-\Omega)\right)^*\right)
 \right)\\
 \Omega_\mathrm{eff}&\approx
 	\Omega_\mathrm{m}
	+  x_\mathrm{zpf}^2 G^2 \mathrm{Im} \left(
 	|\bar a_+|^2 \left(L_+(\bar\Delta+ \Omega)- \left(L_+(\bar\Delta-\Omega)\right)^*\right)+
  	|\bar a_-|^2  \left(L_-(\bar\Delta+ \Omega)- \left(L_-(\bar\Delta-\Omega)\right)^*\right)
 \right).
\end{align}

\section{Calculation of the imprecision-backaction product}
In the context of quantum measurements \cite{SIBraginsky1992}, it is interesting to characterize the sources of noise responsible for the mechanical displacement measurement uncertainty.
For a given mechanical spectra, the measured (double-sided, symmetrized) spectral density of displacement fluctuations is given by
\begin{equation}
 S_\mathrm{xx}^\mathrm{meas}(\Omega) = S_\mathrm{xx}^\mathrm{imp}(\Omega) + | \chi_\mathrm{eff}(\Omega) |^{2} S_\mathrm{FF}(\Omega)
\end{equation}
where $S_\mathrm{xx}^\mathrm{imp}(\Omega)$ describes the measurement imprecision due to apparent displacement fluctuations, which are actually caused by noise in the displacement transducer itself. 
$S_\mathrm{FF}(\Omega)$ is the force noise acting on the mechanical oscillator, and $\chi_\mathrm{eff}(\Omega)$ its effective mechanical susceptibility.
It is particularly interesting to evaluate these quantities for the lowest occupancy obtained at the optimum detuning of $\bar{\Delta} = - \Omega_\mathrm{m} - \frac{\gamma}{2}$ and at the Fourier frequency $\Omega=\Omega_\mathrm{m}$.

In our experiment, the \textit{measurement imprecision}  is dominated by shot noise, and we extract a value of 
\begin{align*}
S_\mathrm{xx}^\mathrm{imp} \equiv S_\mathrm{xx}^\mathrm{imp}(\Omega_\mathrm{m}) = ( 3.2 \times 10^{-19} \,\unit{m / \sqrt{Hz} } )^{2}
\end{align*}
from the fit to the background of the recorded mechanical spectrum (Fig. $4$ from the main manuscript).
Its measured linear dependence  on the laser input power $P_\mathrm{in}$ shows that it is strongly dominated by the quantum noise of the input laser.
This behavior is indeed expected at the frequencies of interest in our work, where classical quadrature fluctuations are negligible in Ti:sapphire lasers.

The thermal force noise (for $\frac{k_\mathrm{B} T}{\hbar \Omega_\mathrm{m}} \gg 1$) driving the mechanical oscillator is given by
\begin{equation}
S_\mathrm{FF}^\mathrm{the} \equiv S_\mathrm{FF}^\mathrm{the}(\Omega_\mathrm{m}) = 2 m_\mathrm{eff} k_{\mathrm{B}} T \Gamma_{\mathrm{m}}(T)
\end{equation}
given by the fluctuation-dissipation theorem.
In the presence of dynamical backaction, we can estimate this force noise from the more directly measured linewidth $\Gamma_\mathrm{eff}$ and noise temperature $T_{\mathrm{eff}}$ using $T_{\mathrm{eff}}=T \cdot \Gamma_{\mathrm{m}}(T) / \Gamma_{\mathrm{eff}}$, and
\begin{equation}
 S_\mathrm{FF}^\mathrm{the} = 2 m_\mathrm{eff} k_{\mathrm{B}} T_{\mathrm{eff}} \Gamma_{\mathrm{eff}}
\end{equation}
It evaluates to 
\begin{align*}
  S_\mathrm{FF}^\mathrm{the} = ( 8 \pm 2 \times 10^{-15} \,\unit{N / \sqrt{Hz}} )^{2}
\end{align*}
 where $\Gamma_{\mathrm{eff}}$ and $T_{\mathrm{eff}}$ are extracted from the fits to the detuning series, evaluated at the detuning $\bar{\Delta} = - \Omega_\mathrm{m} - \frac{\gamma}{2}$ as described in the main manuscript.
This value gives a conservative estimate of the \textit{classical measurement backaction}, considering effectively \textit{all force noise} present in the system (including thermal noise due to the non-zero cryostat temperature) as a classical backation of the measurement.

A less conservative estimate on the backaction of the actual displacement measurement using the laser coupled to the WGM can be made by separating two different contributions in the force noise,
\begin{equation}
 S_\mathrm{FF}^\mathrm{the} = S_\mathrm{FF}^\mathrm{cryo} + S_\mathrm{FF}^\mathrm{ba},
\end{equation}
where $S_\mathrm{FF}^\mathrm{cryo}$ is the Langevin force noise due to the finite cryostat temperature $T_\mathrm{cryo}$ and $S_\mathrm{FF}^\mathrm{ba}$ the thermal backaction in the form of excess Langevin force noise due to the heating of the cavity by laser light.
$S_\mathrm{FF}^\mathrm{ba}$ gives an estimate of the classical perturbation of the system by the measurement, the \textit{classical excess backaction}, which is technically avoidable.

The thermal force noise originating from the bath
\begin{equation}
S_\mathrm{FF}^\mathrm{cryo} = 2 m_\mathrm{eff} k_{\mathrm{B}} T_{\mathrm{cryo}} \Gamma_{\mathrm{m}}(T_{\mathrm{cryo}})
\end{equation}
is estimated to 
\begin{align*}
S_\mathrm{FF}^\mathrm{cryo} = ( 5 \pm 1 \times 10^{-15} \,\unit{N / \sqrt{Hz}} )^{2}.
\end{align*}
$T_{\mathrm{cryo}}$ and $\Gamma_{\mathrm{m}}(T_{\mathrm{cryo}})$ are extracted from independent low input power measurements where the RBM is thermalized to the cryostat temperature.

Consequently, the excess classical backaction evaluates to 
\begin{align*}
S_\mathrm{FF}^\mathrm{ba} = ( 6 \pm 2 \times 10^{-15} \,\unit{N / \sqrt{Hz}} )^{2} 
\end{align*}
 and accounts for 60 \% of the thermal force fluctuations driving the mechanical oscillator.

In addition to classical backaction, the quantum fluctuations of the intracavity photon number give rise to a \textit{quantum measurement backaction} for which the force noise is given by
\begin{equation}
  S_{\mathrm{FF}}^{\mathrm{qba}} \equiv S_{\mathrm{FF}}^{\mathrm{qba}}(\Omega_{\mathrm{m}}) \approx \frac{2 \hbar G^{2} P_{\mathrm{in}} \etac}{\bar{\omega}_\mathrm{c} \Omega_{\mathrm{m}}^{2}} = \frac{4 g_{0}^{2} m_{\mathrm{eff}} P_{\mathrm{in}} \etac}{\bar{\omega}_\mathrm{c} \Omega_{\mathrm{m}}}
\end{equation}
in the case of high resolved sideband factor $\frac{\Omega_{\mathrm{m}}}{\kappa} \gg 1$ \cite{SISchliesser2009a} and at the detuning of interest.
It is of the order of $( 1 \times 10^{-15} \,\unit{N / \sqrt{Hz}} )^{2}$ in our case, negligible compared to the classical backaction.

Therefore, a conservative estimate of the \textit{imprecision-backaction product} is given by (for $\frac{k_\mathrm{B} T}{\hbar \Omega_\mathrm{m}} \gg 1$) 
\begin{align*}
\sqrt{S_\mathrm{xx}^\mathrm{imp} ( S_\mathrm{FF}^\mathrm{the} + S_\mathrm{FF}^\mathrm{qba})} \approx \sqrt{S_\mathrm{xx}^\mathrm{imp} S_\mathrm{FF}^\mathrm{the}} = ( 49 \pm 8 )\frac{\hbar}{2}.
\end{align*}
In an ideal quantum measurement \cite{SIBraginsky1992}, this product is equal to $\frac{\hbar}{2}$, corresponding to the optimal compromise between quantum imprecision and quantum backaction, both arising from the quantum fluctuations of the optical field quadratures.

As it is shown in the main manuscript, laser absorption heating, responsible for the classical excess backaction $S_{\mathrm{FF}}^{\mathrm{ba}}$, is mainly caused by scattered light off the tapered fiber being absorbed by the toroid (in our case by dust particles on the tapered fiber originating from particles in the air of our laboratory).
It is thus within technical reach to strongly reduce this effect and perform measurements where light induced backaction would be dominated by quantum fluctuations alone.
%


%


\end{document}